\RequirePackage{fix-cm}
\documentclass[smallextended]{svjour3}       
\smartqed  
\usepackage{graphicx}
\begin{document}

\title{Topics in Low-Energy QCD with Strange Quarks
}


\author{Wolfram Weise         
}


\institute{W. Weise \at
Physik-Department, Technische Universit\"at M\"unchen, D-85747 Garching, Germany\\
and\\ECT*, Villa Tambosi, I-38123 Villazzano (TN), Italy\\ 
              \email{weise@tum.de}          
}

\date{Received: date / Accepted: date}

\maketitle

\begin{abstract}
Recent developments and pending issues in low-energy strong interaction physics
with strangeness are summarized. Chiral SU(3) effective field theory has progressed as the appropriate theoretical framework
applied to antikaon- and hyperon-nuclear systems. Topics include antikaon-nucleon interactions and the $\Lambda(1405)$, $\bar{K}NN$ systems, recent developments in hyperon-nucleon interactions and strangeness in dense baryonic matter with emphasis on new constraints from neutron star observations. 
\end{abstract}

\section{Introduction}
\label{intro}
Within the hierarchy of quark masses in QCD, strange quarks are special. With their mass $m_s \sim 0.1$ GeV, $s$-quarks are not extremely light (unlike the $u$ and $d$ quarks with their masses $m_{u,d}$ of only a few MeV), but at the same time still well separated on the mass scale from the heavy ($c$, $b$ and $t$) quarks. Systems with strange quarks are thus of interest as test areas for the interplay between spontaneous and explicit chiral symmetry breaking in low-energy QCD.

 High-precision antikaon-nucleon threshold physics is one such testing ground. The driving attractive s-wave $K^-N$ interactions at threshold are determined entirely by the kaon energy and decay constant. The leading-order threshold amplitudes are $T(K^-p)|_{thr} =  2\,T(K^-n)|_{thr} = m_K/f^2$. The appearance of the pseudoscalar decay constant $f \sim 0.1$ GeV is characteristic of spontaneously broken chiral symmetry in low-energy QCD, for which $f$ is an order parameter\footnote{In the chiral limit, $f \simeq 86$ MeV. Explicit chiral symmetry breaking shifts this value to the empirical $f_\pi = 92.2\pm 0.2$ MeV for the charged pion and $f_K = 110.4\pm0.8$ MeV for the charged kaon \cite{PDG2012}.}. The kaon mass $m_K$ (actually the kaon energy at zero momentum) reflects explicit breaking of chiral symmetry by the non-vanishing (strange) quark mass, $m_K^2$ being proportional to $m_s + m_u$. In the chiral limit of vanishing quark masses the corresponding threshold amplitude for the interaction of a pseudoscalar Nambu-Goldstone boson with a nucleon would vanish in accordance with Goldstone's theorem. 

Confinement implies that QCD in the low-energy limit is realized as a theory of hadronic rather than quark-gluon degrees of freedom. Spontaneous chiral symmetry breaking implies further that the appropriate framework is Chiral Effective Field Theory (ChEFT), a systematic approach describing the interactions of the pseudoscalar Nambu-Goldstone bosons amongst each other and with ``heavy" sources such as baryons. For the two-flavor case with almost massless $u, d$ quarks and spontaneously broken $SU(2)_L\times SU(2)_R$ symmetry, chiral perturbation theory (ChPT) has been applied successfully to $\pi\pi$, $\pi N$ and $NN$ scattering. Moreover, in-medium ChPT has been developed as an appropriate tool for treating the nuclear many-body problem at moderate densities (for recent reviews see \cite{HKW2013,HRW2014} and refs. therein). 

Three-flavor QCD at low energy is represented by $SU(3)_L\times SU(3)_R$ ChEFT where now the kaon mass appears prominently in the symmetry-breaking mass term. This theory involves the pseudoscalar meson octet coupled to the baryon octet. However, perturbative (ChPT) methods are in general not applicable in the sector with strangeness. This is partly a consequence of the stronger explicit chiral symmetry breaking by the strange quark mass which implies, in particular, significantly stronger interactions of antikaons as compared to those of pions near their respective thresholds. For example, the existence of the $\Lambda(1405)$ just 27 MeV below $\bar{K}N$ threshold rules out a perturbative ChPT treatment as an expansion in powers of ``small" momenta. Non-perturbative strategies, based on the chiral $SU(3)$ EFT effective Lagrangian at next-to-leading order as input but solving coupled-channels Lippmann-Schwinger equations to all orders, have been well established as the method of choice to deal with these problems \cite{KSW1995}. While sacrificing the rigorous power-counting scheme of ChPT, the gain in physics insights using this non-perturbative approach emerges as a major benefit.

In the following, selected topics of recent interest in this field are discussed, including low-energy $\bar{K}N$ and $\bar{K}NN$ interactions, progress in understanding the nature and structure of the $\Lambda(1405)$, new developments concerning hyperon-nucleon interactions derived from ChEFT, and the possible role of strangeness in dense baryonic matter in view of the established existence of two-solar-mass neutron stars.    

\section{Low-energy $\bar{K}$-nucleon interaction and structure of the $\Lambda(1405)$}
\label{sec:1}
\begin{figure}
\includegraphics[width=1.0\textwidth]{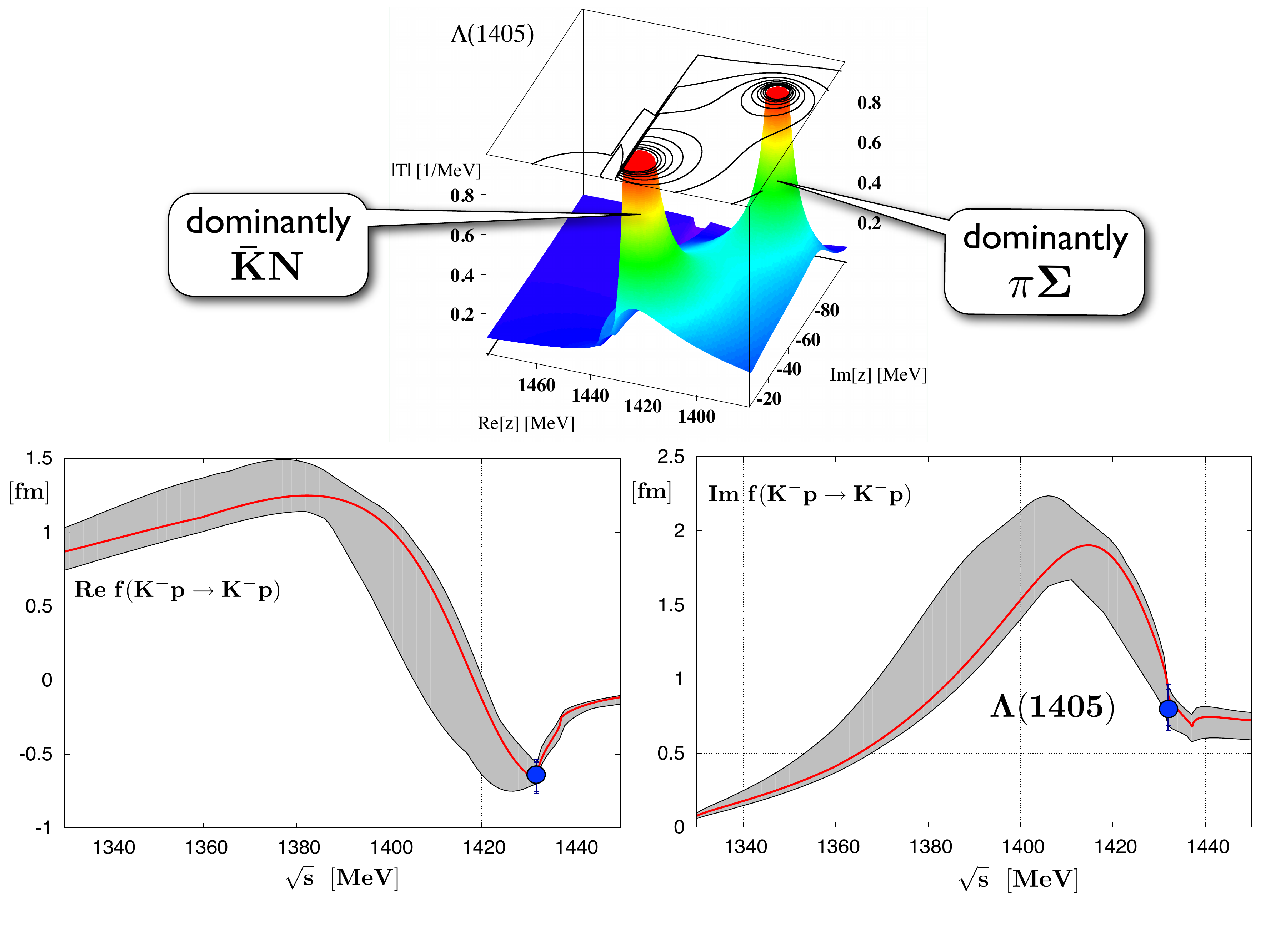}
\caption{Lower panel: Real and imaginary parts of the $K^-p$ scattering amplitude from chiral SU(3) coupled-channels dynamics \cite{IHW2012}. The threshold points indicating the real and imaginary parts of the corresponding scattering length are constrained by the SIDDHARTA kaonic hydrogen measurement \cite{Siddharta2011}. Upper panel: two-poles scenario of coupled $\bar{K}N$ and $\pi\Sigma$ channels \cite{HJ2012}.}
\label{fig:1} 
\end{figure}

Applications of chiral SU(3) coupled-channels dynamics to threshold and substhreshold $\bar{K}N$ interactions have focused prominently on the $K^-p$ system. Its dynamics includes the  $K^-p \leftrightarrow \bar{K}^0 n$ charge exchange channel and the strong coupling to the $\pi\Sigma$ continuum. Goldstone's theorem implies that the driving attractive s-wave interactions are proportional to the energies of the participating pseudoscalar mesons. With input constrained by the strong interaction energy shift and width deduced from the SIDDHARTA kaonic hydrogen measurement \cite{Siddharta2011}, calculated real and imaginary parts of the $K^-p$ amplitude \cite{IHW2012} are shown in Fig. \ref{fig:1}. The resulting complex $K^-p$ scattering length (including Coulomb corrections) is \cite{IHW2012}:
\begin{eqnarray}
\mathrm{Re}\,a(K^-p) = (-0.65 \pm 0.10) \,\mathrm{fm}~, ~~~\mathrm{Im}\,a(K^-p) = (0.81 \pm 0.15)\,\mathrm{fm}~.
\label{eq:akp}
\end{eqnarray}
The uncertainties in $a(K^-p)$ derive primarily from those of the kaonic hydrogen data. Further
extensions of such calculations based on chiral $SU(3)$ effective field theory have recently been reported in \cite{GO2013,MM2013}.

The $SU(3)$ ChEFT coupled-channels approach also predicts $K^-n$ amplitudes. While the weaker attraction in this $I = 1$ channel does not produce a quasibound state or resonance, 
the NLO calculations still suggest a sizeable $K^-n$ scattering length \cite{IHW2012}: 
\begin{eqnarray}
\mathrm{Re}\,a(K^-n) = (0.57^{+0.04}_{-0.21} )\,\mathrm{fm}~, ~~~\mathrm{Im}\,a(K^-n) = (0.72^{+0.26}_{-0.41})\,\mathrm{fm}~.
\label{eq:akn}
\end{eqnarray}
It is important to  provide empirical constraints for this quantity in order to arrive at complete information for both isospin $I = 0$ and $I = 1$ channels of the $\bar{K}N$ system. 
This can be achieved by accurate antikaon-deuteron threshold measurements such as the
planned SIDDHARTA-2 kaonic deuterium experiment and a related proposal at J-PARC.

The $\Lambda(1405)$ with isospin $I = 0$ emerges from coupled-channels dynamics as a quasibound $\bar{K}N$ state embedded in the $\pi\Sigma$ continuum. A characteristic feature of the chiral $SU(3)$ coupled-channels approach is the appearance of two poles in the T matrix (see \cite{HJ2012} for a review and refs. therein). The calculation that leads to the amplitude in Fig. \ref{fig:1} produces a pole located at $(E_1, \Gamma_1/2) = (1424\pm15,~26\pm5)$ MeV and a second one at $(E_2,\Gamma_2/2) = (1381\pm15,~81\pm8)$ MeV in the complex energy plane. The first pole has a dominant $\bar{K}N$ bound state component. 
This pole is quite well determined, with its imaginary part representing the decay width into the open $\pi\Sigma$ phase space driven by the $\bar{K}N\rightarrow \pi\Sigma$ interaction. The location of the second pole is more ambiguous, less well determined by the fits to existing scattering data and kaonic hydrogen measurements.  Its dominant component reflects a broad resonance structure in the $\pi\Sigma$ channel. The coexistence of these two coupled modes implies that the $\Lambda(1405)$ is not described by a single, unique spectral function but by an entanglement of the two modes. The $\pi\Sigma$ spectra observed in different photon- and hadron-induced reactions are expected to differ in their shape and location of their maximum, depending on which of the $\bar{K}N$ or $\pi\Sigma$ components of the coupled modes are more strongly involved in the particular reaction mechanism at work. For example, the maximum of the $\pi^-\Sigma^+$ invariant mass spectrum produced in $\gamma p\rightarrow K^+\pi^- \Sigma^+$ photoproduction at JLab \cite{CLAS2014} is observed around 1420 MeV,
as expected for a process driven by a primary $\gamma \rightarrow K^+K^-$ vertex and followed by the t-channel exchange of the $K^-$ that is absorbed by the proton and then converted to the observed $\pi^-\Sigma^+$ final state. On the other hand, the more complex hadron-induced reactions tend to emphasize more strongly the primary $\pi\Sigma$ channels and produce a maximum in the $\pi\Sigma$ spectrum around the nominal 1405 MeV.

Hence the $\Lambda(1405)$ is obviously not a standard quark model baryon. Consider a Fock space expansion
\begin{eqnarray}
|\Lambda^*\rangle = a\,|uds\rangle + b\,|(udu)(\bar{u}s)\rangle + c\,|(uus)(\bar{u}d)\rangle + \dots~~,
\end{eqnarray}
where the first term on the r.h.s. corresponds to a ``bare" three-quark state while the subsequent terms
schematically describe $N\bar{K}$ and $\Sigma\pi$ quasimolecular configurations. Actually any baryon state has an expansion of this kind. The issue is how the coefficients of this Fock expansion arrange themselves, governed by the strong interactions of the constituents.  For a baryonic state in the proximity of a meson-baryon threshold, a quasi-molecular structure tends to be favored. 

\begin{figure*}
\includegraphics[width=0.65\textwidth]{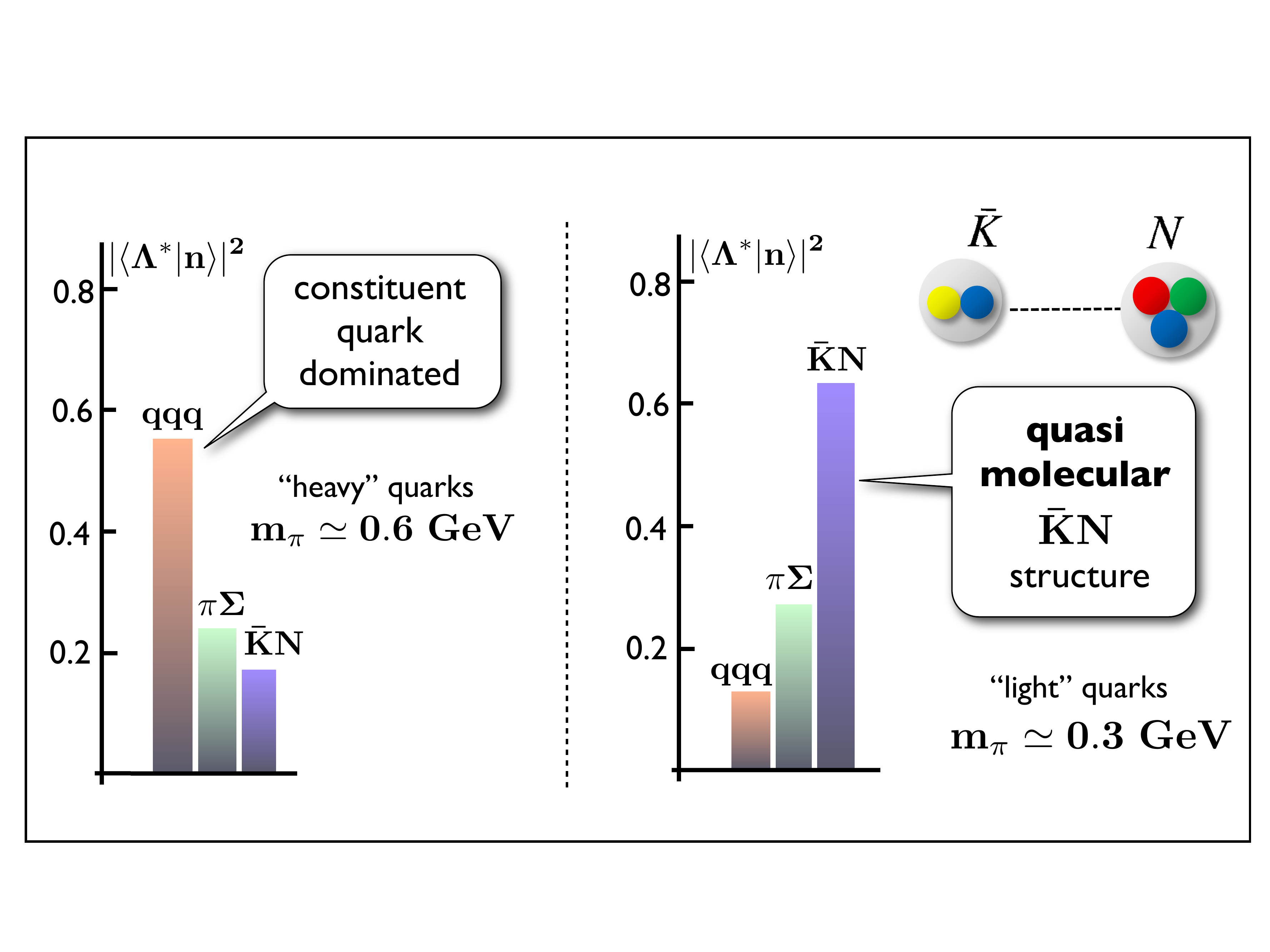}
\caption{Leading components of the $\Lambda(1405)$ deduced from a lattice QCD
computation \cite{Hall2014} for different light-quark masses indicated in terms of the
pion mass. }
\label{fig:2}      
\end{figure*}

In this context a recent lattice QCD study \cite{Hall2014} of the $\Lambda(1405)$ comes to interesting conclusions (see Fig.\ref{fig:2}).
When all quarks are kept at a relatively large mass scale characteristic of the strange quark mass, the $\Lambda(1405)$ behaves more like a three-quark system reminiscent of the naive quark model. As the light quarks approach their physical masses, the $\Lambda(1405)$ emerges as a state with dominant quasi-molecular $\bar{K}N$ structure. This is where lattice QCD meets chiral $SU(3)$ coupled-channels dynamics.   

The $\Lambda(1405)$ thus figures as prototype of a non-conventional hadronic species
displaying a quasi-molecular weak-binding structure close to a threshold (in this case the $\bar{K}N$ threshold). Several cases of analogous phenomena are observed and discussed in the physics with charmed quarks close to thresholds involving $D$ or $D^*$ mesons. 

\section{Brief report on $\bar{K}NN$ three-body systems}
\begin{figure*}
\includegraphics[width=0.6\textwidth]{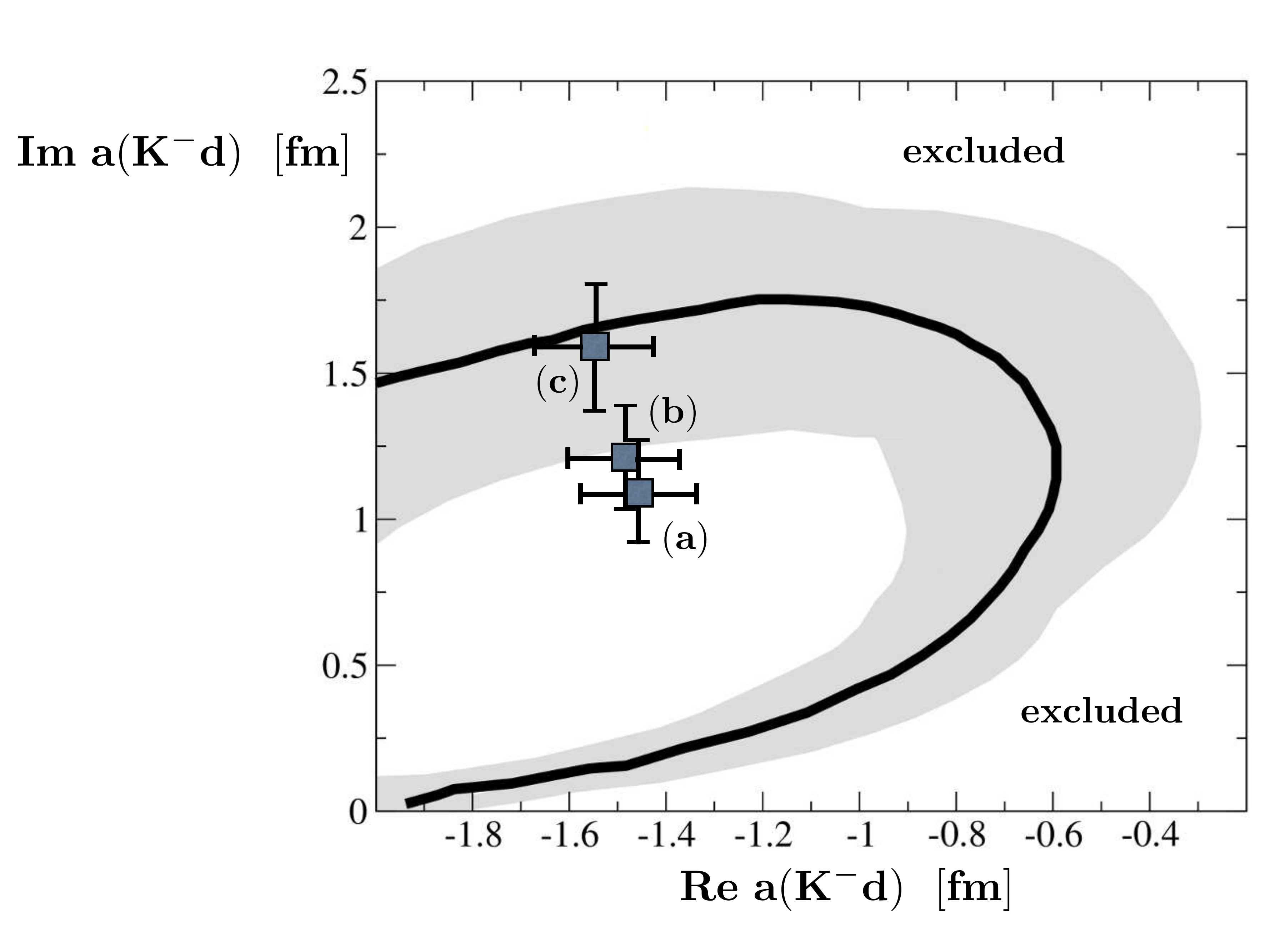}
\caption{Real and imaginary parts of the $K^-$ deuteron scattering length subject to empirical constraints from $K^-p$ data. (a): non-relativistic EFT calculation \cite{DM2011}; (b): Faddeev calculation using separable amplitudes \cite{Sh2012}; (c) three-body multiple scattering calculation \cite{Ohnishi2014} using $K^-p$ and $K^-n$ amplitudes from \cite{IHW2012}. }
\label{fig:3}      
\end{figure*}

Low-energy antikaon-neutron interactions are accessible through the investigation of the ${\bar{K}}$-deuteron three-body system. Examples of calculations of the $K^-$ deuteron scattering length using data-constrained $K^-p$ input together with predicted $K^-n$ amplitudes are presented in Fig.\,\ref{fig:3}. Different approaches (non-relativistic EFT \cite{DM2011}, Faddeev \cite{Sh2012} and three-body multiple scattering calculations \cite{Ohnishi2014}) are seen to arrive at reasonably consistent results. Starting from the $K^-p$ and $K^-n$ amplitudes derived in \cite{IHW2012} gives \cite{Ohnishi2014}:
\begin{eqnarray}
a(K^-d) = (-1.55 +\mathrm{i}\,1.66)\,\mathrm{fm}~~, 
\label{eq:akd}
\end{eqnarray}
with 10-20 \% uncertainties \cite{Gal2007} resulting from the ``fixed scatterer" (no-recoil) treatment of the nucleons, uncertainties in the $K^-n$ amplitude and neglect of  $K^-d\rightarrow YN$ absorption. Note that while the real parts of the individual scattering lengths (\ref{eq:akp}) and (\ref{eq:akn}) would tend to cancel in leading order, the $K^-pn \leftrightarrow \bar{K}^0nn$ charge exchange process is important in determining the result (\ref{eq:akd}).

Predictions for $a(K^-d)$ can in principle be tested by measuring the strong-interaction energy shift $\Delta E$ and width $\Gamma$ of 1s kaonic deuterium. This requires a non-trivial three-body calculation of the $K^-$ deuteron atomic system in the presence of strong interactions, but a good estimate can already be obtained using the improved Deser formula \cite{MRR2004}:
\begin{eqnarray}
\Delta E - {\mathrm{i}\Gamma\over 2} = - \,{2\mu^2\alpha^3\,a(K^-d)\over 1-2\mu\alpha(1-\ln\alpha)\,a(K^-d)}~~, 
\end{eqnarray}
where $\mu$ is the kaon-deuteron reduced mass and $\alpha$ is the fine structure constant. 
With $a(K^-d)$ taken from (\ref{eq:akd}) this gives $\Delta E = 0.87$ keV and $\Gamma = 1.19$ keV, again with an estimated uncertainty of 10 - 20 \%.

A great amount of activities has been focused in recent times on the possible existence of quasibound antikaon-nuclear clusters, inspired by earlier phenomenological considerations \cite{AY}. A prototype example is the $K^-pp$ system. Here we briefly report on theoretical expectations using interactions that are based on chiral $SU(3)$ dynamics with coupled channels. The present situation is summarized in Table \ref{tab:1} which collects results from three different variational and Faddeev three-body calculations. While these computations differ in details, there is now a high degree of at least qualitative consistency in the predictions that a possible $K^-pp$ cluster is expected to be weakly bound (with a binding energy not exceeding about 15-20 MeV), but short-lived with a width around 40 MeV or larger. Such a large width would make experimental searches for a quasibound $K^-pp$ state not an easy task. 

\begin{table}
\caption{Binding energies and widths of the $K^-pp$ system from variational \cite{DHW2009,BGL2012}
and Faddeev \cite{IKS2010} calculations using energy-dependent chiral SU(3) based interactions as input. }
\label{tab:1}       
\begin{tabular}{cccc}
\hline\noalign{\smallskip}
 & Variational \cite{DHW2009} & Faddeev \cite{IKS2010} & Variational \cite{BGL2012} \\
\noalign{\smallskip}\hline\noalign{\smallskip}
B\,[MeV] & 17-23 & 9-16 & 16\\
$\Gamma$\,[MeV] & 40-70 & 34-46 & 41\\
\noalign{\smallskip}\hline
\end{tabular}
\end{table}

The weak-binding scenario resulting from chiral $SU(3)$ dynamics with energy-dependent
interactions and strong $\bar{K}N \leftrightarrow \pi\Sigma$ channel coupling differs conceptually and quantitatively from the earlier phenomenological picture \cite{AY} 
that assumes an energy-independent local $\bar{K}N$ potential and generates the $\Lambda(1405)$ as a single pole in the $\bar{K}N$ amplitude. While both approaches (chiral $SU(3)$ dynamics vs. phenomenology) reproduce kaonic hydrogen and the existing (admittedly not very accurate) $K^-p$ scattering data, they differ significantly in their extensions below $\bar{K}N$ threshold. The energy-independent potential model would suggest a much more strongly bound $K^-pp$ state than the approach based on chiral $SU(3)$ dynamics. 

A possible way of distinguishing empirically between chiral $SU(3)$ two-pole dynamics and single-pole phenomenolgy is offered by the $K^-d\rightarrow n\pi\Sigma$ reaction with an incident $K^-$ momentum around 1 GeV/c and detection of the outgoing neutron, as discussed in \cite{OIHHW2014}. The neutron energy spectrum reflects the $\pi\Sigma$ mass distribution. Below $K^-p$ threshold the spectra of the three charge combinations, $\pi^-\Sigma^+$, $\pi^0\Sigma^0$ and $\pi^+\Sigma^-$, are expected to feature characteristic differences between the energy-dependent two-poles approach and the energy-independent single-pole phenomenology. A corresponding experimental proposal (E31) is on the way at J-PARC.

Several dedicated experiments are now being pursued in a focused search for a possible $K^-pp$ quasibound state. The HADES collaboration \cite{HADES2014} has performed a detailed partial wave analysis of the $pp\rightarrow K^+\Lambda p$ reaction with 3.5 GeV incident protons. A possible quasibound $K^-pp$ cluster should show up via its decay into $\Lambda p$. No significant signal was found, and an upper limit for the cluster production cross section of about 1 - 4 $\mu$b has been deduced depending on the assumed width of the hypothetical $K^-pp$ quasibound state. Further recent examples of dedicated activities are the $^3$He$(K^-,n)X$ (E15) \cite{Hash2014} and $d(\pi^+,K^+)X$ (E27) \cite{Ichi2014-1,Ichi2014-2} experiments at J-PARC. The first set of E15 data \cite{Hash2014} shows no structure in the deeply-bound region but a small excess below $K^-pp$ threshold in the energy range expected for a possible weakly-bound cluster. On the other hand, E27 observes a very broad structure in a range 60 - 140 MeV below $K^-pp$ threshold \cite{Ichi2014-2}. While definitive conclusion cannot be drawn so far, the more focused exclusive measurements and analysis of the final state $X$ continue. 

\section{Hyperon-nucleon interactions and strangeness in dense matter}
Along with successful applications to kaon- and antikaon-baryon systems, chiral $SU(3)$ effective field theory provides also the appropriate framework for a theory of hyperon-nucleon interactions.
Recent progress has been made \cite{Haid2013} in developing this ChEFT framework at next-to-leading order (NLO) such that the relevant one- and two-meson exchange mechanisms involving the complete pseudoscalar meson octet are properly incorporated. The important two-pion exchange $\Lambda N$ interaction with $\Sigma N$ intermediate states emerges naturally in this approach.

\begin{figure*}
\includegraphics[width=0.7\textwidth]{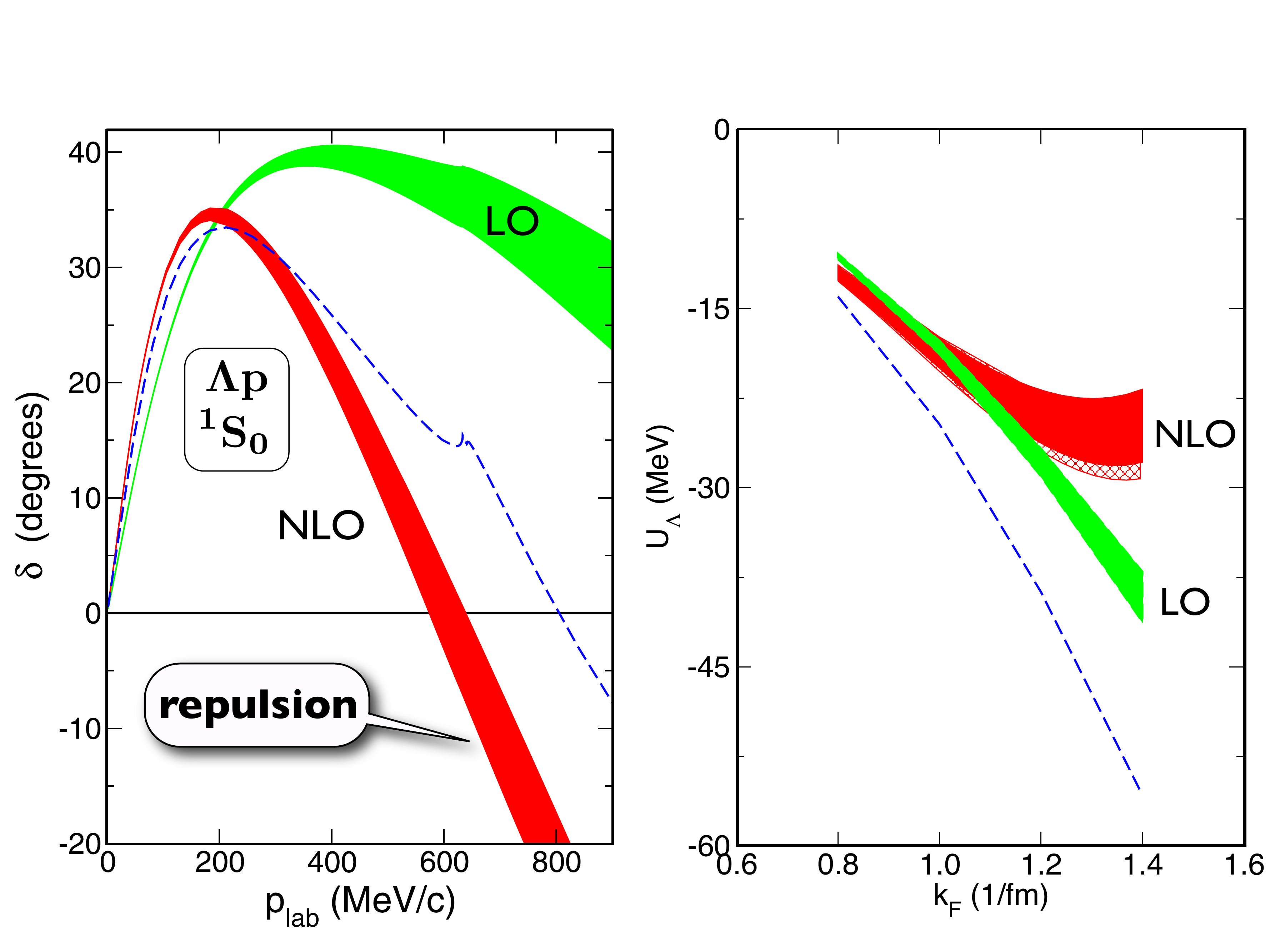}
\caption{Left: Singlet s-wave $\Lambda p$ phase shift derived from the haperon-nucleon interaction based on chiral $SU(3)$ EFT \cite{Haid2013} in leading (LO) and next-to-leading order (NLO). Uncertainty bands refer to variations within a range of momentum space cutoffs 0.5 - 0.7 GeV. The dashed curve is the result obtained with the phenomenological J\"ulich04 potential. Right: Single particle potential $U_\Lambda$ of a $\Lambda$ hyperon in nuclear matter as function of Fermi momentum \cite{HM2014}. Notations as in the left figure. }
\label{fig:4}      
\end{figure*}

Fig.\,\ref{fig:4} (left) shows as an example the momentum dependence of the $^1$S$_0$ phase shift for $\Lambda p$ scattering calculated in such a framework. While the leading-order (LO) result displays attraction over the whole range of momenta, the NLO calculation shows a turnover from attraction at low momenta to repulsion at high momenta. This behavior is qualitatively consistent with characteristic features of hyperon-nucleon potentials deduced from lattice QCD \cite{Aoki2012} where it is found that intermediate-range attraction turns to strong short-range repulsion at distances $r \simeq 0.5$ fm. The $\Lambda$ single particle potential $U_\Lambda$ in nuclear matter derived from a Brueckner G-matrix calculation using the ChEFT interaction (with inclusion of the dominant $^3$S$_1$-$^3$D$_1$ channel) is shown in Fig.\,\ref{fig:4} (right). The dependence of $U_\Lambda$ on the nuclear Fermi momentum $k_F$ demonstrates the stabilization of the NLO potential at values consistent with properties of $\Lambda$ hypernuclei. Repulsive effects are expected to act more prominently at higher densities, several times the density of normal nuclear matter ($\varrho_0 \simeq 0.15$ fm$^{-3}$).

\begin{figure*}
\includegraphics[width=0.65\textwidth]{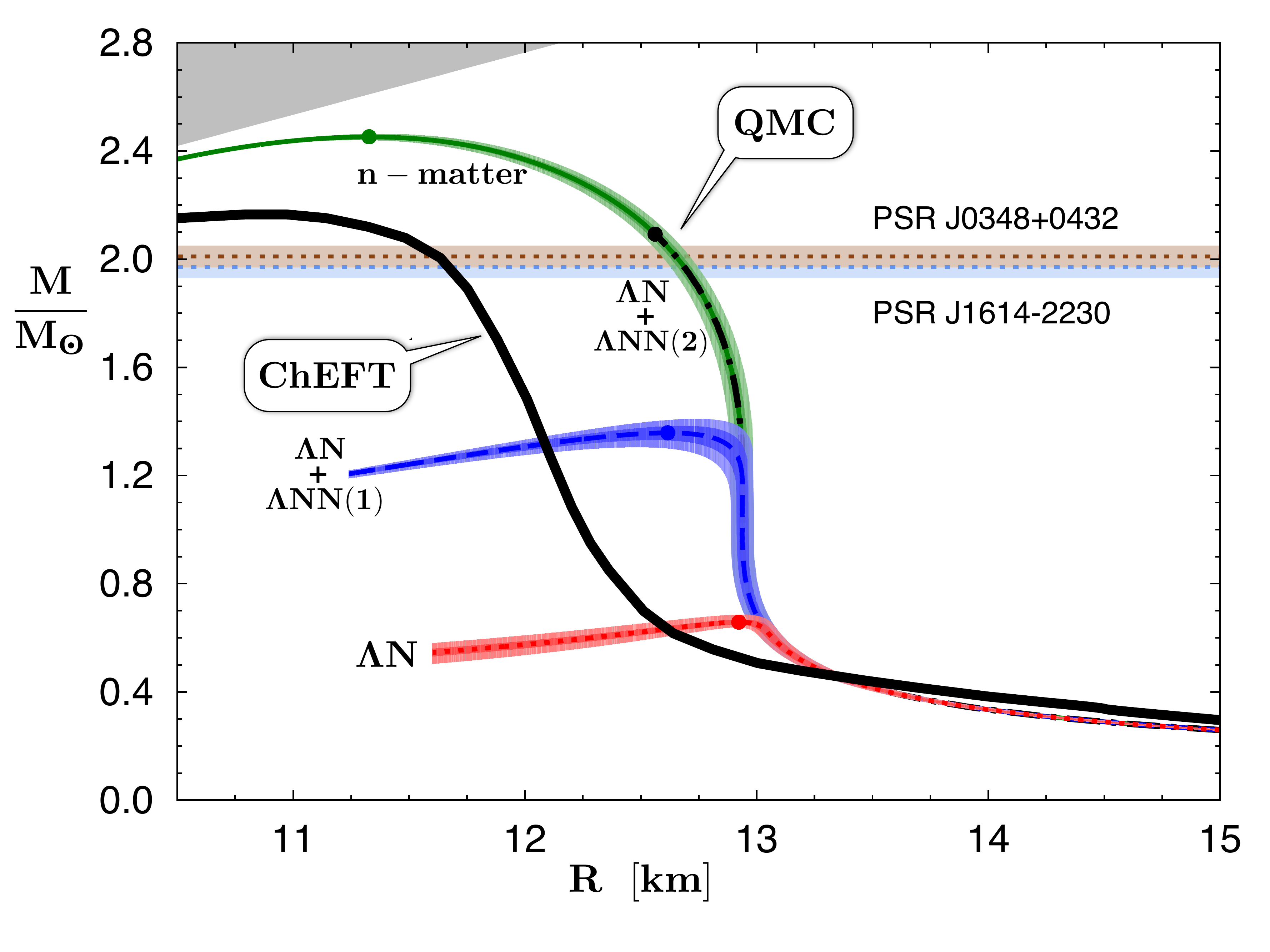}
\caption{Mass-radius relations for neutron stars. Solid black curve: ChEFT result (nucleon + pion degrees of freedom) taken from \cite{HW2014}; colored curves: QMC computations \cite{LLGP2014} including $\Lambda$ hyperons with phenomenological $\Lambda N$ forces and two versions of repulsive $\Lambda NN$ three-body interactions. Version $\Lambda NN(2)$ reproduces the systematics of hypernuclear binding energies.}
\label{fig:5}      
\end{figure*}

Such repulsive interactions of the $\Lambda$ with nucleons in dense baryonic matter are presently under active discussion in view of the new constraints implied by the existence of massive neutron stars. The recent observation of two neutron stars with $M \simeq 2M_\odot$ sets strong constraints on the required stiffness of the equation of state (EoS) in order to support such objects against gravitational collapse. An EoS based on ChEFT with ``conventional" nucleon and pion degrees of freedom can produce sufficient pressure at high density, generated by repusive three-body forces and the impact of the Pauli principle on the in-medium nucleon-nucleon effective interaction \cite{HW2014} (see Fig.\,\ref{fig:5}). However,  neutrons in the core of the star tend to be replaced by $\Lambda$ hyperons at densities (typically around 2-3 $\varrho_0$) where this becomes energetically favorable. Then the EoS would soften too much so that maximum neutron star masses of $2M_\odot$ cannot be sustained any more.  

A recent advanced quantum Monte Carlo (QMC) computation of neutron star matter, with hyperons added \cite{LLGP2014}, emphasizes this issue. While this calculation still uses phenomenological $\Lambda N$ input interactions, the conclusions are nonetheless instructive. When parametrized repulsive $\Lambda NN$ three-body forces are added subject to the condition that the systematics of hypernuclear binding energies be reproduced, the admixture of $\Lambda$'s in neutron star matter gets strongly reduced such that the pressure to support a $2M_\odot$ star can be maintained as demonstrated in Fig.\,\ref{fig:5}. The pending question is whether the necessary repulsive effect can be entirely relegated to a hypothetical $\Lambda NN$ three-body force, or whether at least a large part of it comes from momentum-dependent $\Lambda N$ two-body interactions as they appear in the $SU(3)$ ChEFT treatment \cite{Haid2013} at next-to-leading order.

\section{Concluding remarks and summary}
Progress has been made in establishing chiral $SU(3)$ effective field theory as the adequate realization of low-energy QCD with strange quarks. It defines a consistent and well organized coupled-channels framework for kaon-, antikaon- and hyperon-nuclear interactions. The investigation of strangeness $S = -1$ systems with baryon number $B = 1,2$ has reached a more detailed understanding of the $\Lambda(1405)$ as a weakly bound (quasi-molecular) $\bar{K}N$ state imbedded in the strongly coupled $\pi\Sigma$ continuum. Threshold and subthreshold $\bar{K}NN$ physics is proceeding towards a focused experimental program. Concerning the role of strangeness in dense baryonic matter, new constraints imposed by the existence of two-solar-mass neutron stars and the required stiffness of the equation of state imply a quest for strong short-distance repulsion in hypernuclear two- and three-body interactions. Lattice QCD studies are on their way to provide further basic information on these issues.  

\begin{acknowledgements}
Collaborations with Thomas Hell, Tetsuo Hyodo, Yoichi Ikeda and Shota Ohnishi on topics reported in this paper are gratefully acknowledged. Useful comments and suggestions by Avraham Gal are much appreciated. I also thank Daniel Gazda, Norbert Kaiser, Maxim Mai and Stefan Petschauer for discussions. This work was supported in part by BMBF and DFG (CRC 110 ``Symmetries and Emergence of Structure in QCD").
\end{acknowledgements}

\end{document}